\documentclass[prl,aps,reprint,a4paper,superscriptaddress,longbibliography]{revtex4-2}
\usepackage{graphicx}
\usepackage{dcolumn,amsmath}
\usepackage{bm,color}
\usepackage[colorlinks=true,linkcolor=blue,citecolor=blue,urlcolor=blue]{hyperref}

\begin{document}
	

\title{Experimental Certification of Quantum Measurements with Maximally Mixed States}


\author{Jia-He Liang}

\author{Ze-Yan Hao}

\author{Jia-Kun Li}

\affiliation{Laboratory of Quantum Information, University of Science and Technology of China, Hefei 230026, China}
\affiliation{Anhui Province Key Laboratory of Quantum Network, University of Science and Technology of China, Hefei 230026, China}
\affiliation{CAS Center for Excellence in Quantum Information and Quantum Physics, University of Science and Technology of China, Hefei 230026, China}

\author{Kai Sun}

\email{ksun678@ustc.edu.cn}
\affiliation{Laboratory of Quantum Information, University of Science and Technology of China, Hefei 230026, China}
\affiliation{Anhui Province Key Laboratory of Quantum Network, University of Science and Technology of China, Hefei 230026, China}
\affiliation{CAS Center for Excellence in Quantum Information and Quantum Physics, University of Science and Technology of China, Hefei 230026, China}
\affiliation{Hefei National Laboratory, University of Science and Technology of China, Hefei 230088, China}

\author{Zhen-Peng Xu}
\email{zhen-peng.xu@ahu.edu.cn}

\affiliation{School of Physics and Optoelectronics Engineering, Anhui University, 230601 Hefei, China}

\author{Jin-Shi Xu}
\email{jsxu@ustc.edu.cn}

\author{Chuan-Feng Li}
\email{cfli@ustc.edu.cn}

\author{Guang-Can Guo}

\affiliation{Laboratory of Quantum Information, University of Science and Technology of China, Hefei 230026, China}
\affiliation{Anhui Province Key Laboratory of Quantum Network, University of Science and Technology of China, Hefei 230026, China}
\affiliation{CAS Center for Excellence in Quantum Information and Quantum Physics, University of Science and Technology of China, Hefei 230026, China}
\affiliation{Hefei National Laboratory, University of Science and Technology of China, Hefei 230088, China}

\author{Ad\'{a}n Cabello}
\email{adan@us.es}
\affiliation{Departamento de Física Aplicada II, Universidad de Sevilla, E-41012 Sevilla, Spain}
\affiliation{Instituto Carlos I de Física Teórica y Computacional, Universidad de Sevilla, E-41012 Sevilla, Spain}

\date{\today}%


\begin{abstract}
So far, certifying quantum devices from their input-output statistics, under minimal assumptions, required the preparation of specific pure quantum states. Recently, Xu {\em et al.}\ [Phys. Rev. Lett. {\bf 132}, 140201 (2024)] have demonstrated that certain sets of quantum observables can be certified using {\em any} state of full rank. However, their method is restricted to ideal conditions. Here, we address this problem and present an experimentally robust method that eliminates the need of preparing states with high fidelity with respect to specific pure states. We demonstrate the feasibility of the method by experimentally certifying photonic devices implementing Peres' set of 24 ququart observables [J.\ Phys.\ A {\bf 24}, L175 (1991)] and Yu and Oh's set of 13 qutrit observables [Phys.\ Rev.\ Lett.\ {\bf 108}, 030402 (2012)], using maximally mixed states as input. This approach offers a crucial advantage for certifying high-dimensional quantum systems, since it works with maximally mixed and thermal states.
\end{abstract}

\maketitle


{\it Introduction.} Self-testing of quantum systems \cite{mayers2003self,vsupic2020self} enables inferring the functioning of uncharacterized devices from the input-output statistics under minimal assumptions, which include the independence between preparation and measurement devices, and the validity of quantum theory. The method leverages specific predictions of quantum theory, wherein certain input-output correlations are unique up to local isometries \cite{mayers2003self} or unitary transformations \cite{XuZhenpeng}.
Common input-output certification methods include Bell self-testing \cite{mayers2003self}, self-testing via noncontextuality inequalities \cite{SelfTestConHu,SelfTestConIrfan}, and device-independent tests for structures of measurement incompatibility \cite{Quintino:2019PRL,chen2021device}. A general limitation common to all these methods is their requirement of specific pure states, or states with high fidelity with respect to a pure one.
This limitation has been recently overcome by Xu {\em et al.} \cite{XuZhenpeng}, who introduced a method that allows the certification of specific sets of uncharacterized quantum observables using {\em any} full-rank state. Hereafter, we will refer to this method as certification with full rank states (CFR). The CFR is based on the uniqueness, modulo unitaries, of some state-independent contextuality (SI-C) sets \cite{Peres24,BBC-21,CEG-18}. Crucially, for these SI-C sets, attaining the maximal value of a corresponding SI-C witness with any full-rank state implies that, under ideal conditions, all states achieve the same maximal value.
 
Robust certification under experimental imperfections can, in principle, be achieved using semidefinite programming (SDP) \cite{XuZhenpeng}. However, unlike in the ideal case, observing that a full-rank state achieves the maximum value of the witness no longer guarantees that any other state will also achieve this maximum. Therefore, to ensure the reliable certification, the SDP must be run over {\em all} states. This requirement renders the approach impractical, as it is impossible to cover the entire state space. A further drawback is that successful certification demands extremely high experimental precision, in that the admitted deviations of the measured witness ${\cal W}$ from its theoretical maximum must be of the order of $10^{-4}$ or less \cite{XuZhenpeng}. Achieving such a low error threshold poses a significant experimental challenge.

\begin{figure}[h]
\includegraphics[width=0.98\linewidth]{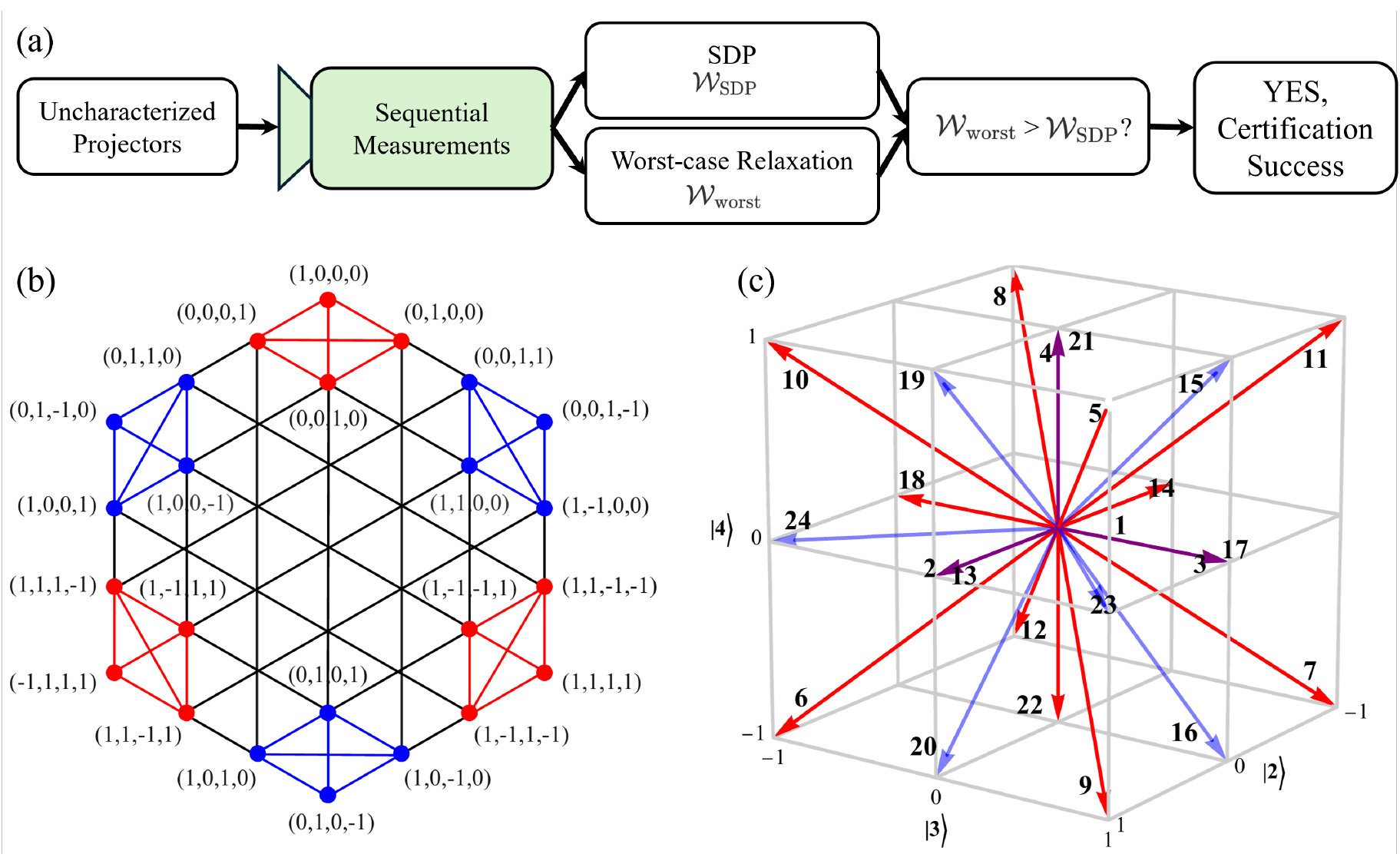}
\caption{(a) Algorithm for certifying uncharacterized projectors using a maximally mixed state. (b)~Orthogonality graph for the Peres-24 set \cite{Peres24}. Vertices represent measurements (represented by rank-one projectors onto the corresponding unnormalized vectors). Vectors in the same straight line or in a clique of the same color are mutually orthogonal. The Hilbert space is spanned by four orthogonal vectors: $|1\rangle, \ |2\rangle, \ |3\rangle, \ |4\rangle$.
(c) Components of the 24 (unnormalized) vectors projected onto the $\{|2\rangle,\ |3\rangle,\ |4\rangle\}$ subspace. A blue arrow indicates the original vector has component $0$ in $|1\rangle$ (i.e., it lies entirely within the subspace). A red arrow indicates it has component $1$ in $|1\rangle$.
Vector $|v_1\rangle=(1,0,0,0)$ is at the origin.} 
\label{fig:process}
\end{figure}

In this Letter, we overcome both challenges. First, we overcome the challenge of testing all quantum states by estimating the value of the SI-C witness for the worst case scenario, $\cal W_{\rm worst}$. For this, we develop a relaxation technique that relies solely on the experimentally observed value for the maximally mixed state. Then, we derive a refined SDP that computes a lower bound, $\cal{W}_{\rm SDP}$, on $\cal{W}_{\rm worst}$ under realistic experimental imperfections. As illustrated in Fig.~\ref{fig:process}(a), whenever $\cal{W}_{\rm worst}> \cal{W}_{\rm SDP}$, we can certify the set of measurements with a precision scaling as the square root of the experimental error. 
Specifically, our SDP incorporates physical and geometrical constraints, enabling us to obtain an experimentally feasible value of $\cal{W}_{\rm SDP}$ under realistic imperfections. This substantially mitigates the requirement of very low experimental errors.

In addition, we demonstrate that the method is truly practical by experimentally certifying two well-known SI-C sets of quantum measurements.
Finally, we discuss why the method offers potential advantages for benchmarking quantum measurements, especially in the case of high-dimensional quantum systems.


{\it Certifying quantum measurements.} A SI-C set $S = \{\Pi_i\}$ can be characterized by its orthogonality graph $G(V,E)$, where the vertex $i$ in the vertex set $V$ represents the projectors $\Pi_i$, and the edge $\{(i,j)\}$ in the edge set $E$ denotes orthogonality relationship $\Pi_i\Pi_j=0$ \cite{SDPCabello}. For example, Fig.~\ref{fig:process}(b) shows the orthogonality graph of a four-dimensional SI-C set called Peres-24 \cite{Peres24,PhysRevLett.134.010201}, with its projection directions illustrated in Fig.~\ref{fig:process}(c). For further details about Peres-24, see \cite{sm}. 

We will use Peres-24 to explain how our method works.
The contextuality of Peres-24 can be revealed using the following SI-C witness \cite{SeqCabello}:
\begin{equation}
	\mathcal W=\sum_{i\in V} P_i - \sum_{i,j\in E} P_{ij},
    \label{eq:witness}
\end{equation}
where $P_i$ and $P_{ij}$ denotes the probability to obtain outcome $1$ from projector $\Pi_i$ and from both $\Pi_i$ and $\Pi_j$, respectively.

Quantum theory \cite{budroni2022kochen} predicts $\mathcal W=6$ for all initial states. We will refer to this value as $\cal W_{\rm opt}$. $\cal W_{\rm opt}$ is also the algebraic maximum constrained by the exclusivity relations \cite{SDPCabello}.

An experimental value for $\mathcal W$ given by \eqref{eq:witness} provides an $(\varepsilon,1/2)$ robust CFR of Peres-24, if, for any projective realization $\{\Pi_i'\}$ of this value under the exclusivity relations encoded in the graph in Fig.~\ref{fig:process}(b), with $\langle \psi|\Pi_i'\Pi_j'\Pi_i'|\psi\rangle \le \varepsilon$, for any state $|\psi\rangle$ there is an isometry ${\cal I}$ such that $|{\cal I}(\Pi_i)-\Pi_i'|\leq {\cal{O}}(\varepsilon^{1/2})$.

An analysis tool based on SDP was proposed~\cite{XuZhenpeng}, which results in a threshold of the witness value such that the robust certification holds when the experimental imperfection is below a given bound $\varepsilon$. 
However, the experimental implementation of this robust method encounters the two significant challenges mentioned above.


\begin{figure*}[t]
	\includegraphics[width=0.8\linewidth]{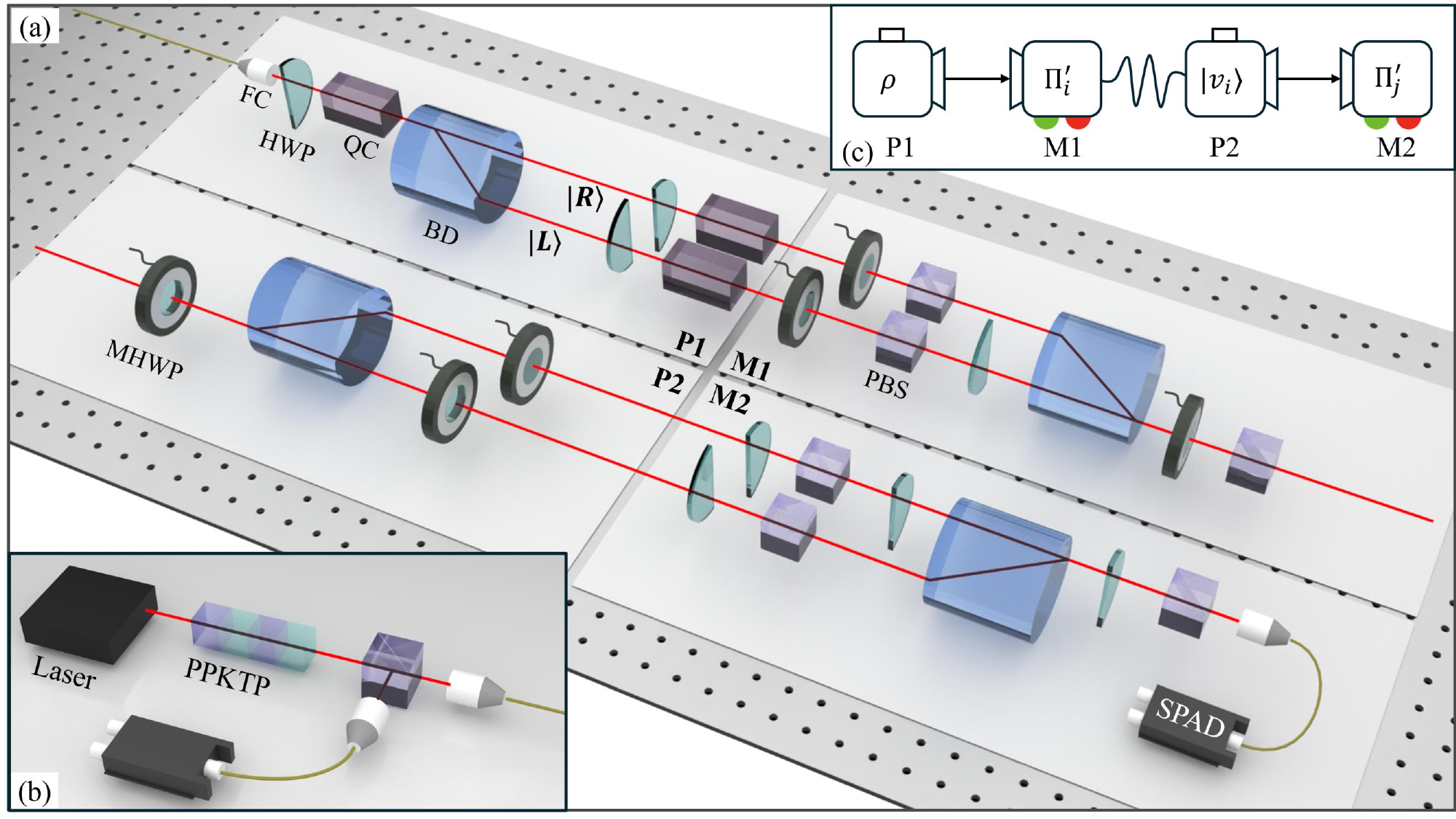}
	\caption{Schematic of experimental setup. (a) Main optical system, (b) 808 nm photon source, (c) logical diagram of the experiment. In (c), P1 and P2 are preparations and M1 and M2 are measurements. Based on the outcome of M1, which measures the projector $\Pi_i'$, the motorized half-wave plates in P2 automatically prepare the corresponding state $|v_i\rangle$, using the method of preparing 4-dimensional state in P1. Abbreviations are
		BD (beam displacer), 
		FC (fiber coupler), 
		HWP (half-wave plate), 
		MHWP (motorized half-wave plate), 
		PBS (polarization beam splitter), 
		PPKTP (periodically polled potassium titanyl phosphate), 
		QC (quartz crystal), and 
		SPAD (single photon avalanche diode).}
	\label{fig:setup}
\end{figure*}


To overcome the challenge of testing all states, we develop a relaxation technique to estimate a lower bound of the minimum witness value $\cal W_{\rm worst}$ for all states, which is derived from the witness value $\mathcal{W}$ of maximally mixed state and the theoretically optimal one $\mathcal W_{\rm opt}$ as follows:
\begin{equation}
	\begin{aligned}
		\mathcal W_{\rm worst} \ge d\,\mathcal{W}-(d-1)\mathcal W_{\rm opt},
	\end{aligned}
	\label{eq:worst}
\end{equation}
where $d$ denotes the dimension of quantum system. As for the evaluation of the experimental value $\mathcal{W}$ according to Eq.~\eqref{eq:witness}, we obtain $P_{ij}$ by $P_i\epsilon_{ij}$ with $\epsilon_{ij} = {\rm tr}(\Pi_i'\Pi_j')$ \cite{XiaoYa,SeqLarsson,SeqCabello}, since $\Pi_i'$ and $\Pi_j'$ are all rank-$1$ in our setup. 

The idea behind Eq.~\eqref{eq:worst} stems from the fact that the maximally mixed state is an even combination of any state together with another $d-1$ states and the worst case can at most absorb all imperfections while other $d-1$ states achieve $\cal W_{\rm opt}$. Then, the linearity of the witness value on the state ensures that the inequality in Eq.~\eqref{eq:worst} holds.
This method significantly reduces the consumption of quantum resources, as it requires experimental data only from the maximally mixed state rather than from an infinite set of states. Further details on $\mathcal W_{\rm worst}$ are discussed in the Supplemental Material \cite{sm}.

In addition, to relax the requirement of extremely low experiment errors, we improve the SDP-based analysis to calculate the witness threshold $\mathcal W_{\rm SDP}$ for $\cal W_{\rm worst}$ for a given imperfect orthogonality $\epsilon_{ij}$. This SDP ensures the non-orthogonality of any two projectors corresponding to two disconnected vertices, and the completeness of each context represented by $d$ fully connected vertices in Fig.~\ref{fig:process}(b). 
Non-orthogonality is verified when minimizing $\epsilon_{ij}$ yields a non-zero optimum for all $(i,j)\notin E$.
For $(m,n)\in E$, $\epsilon_{mn}$ corresponds to a component of positive semi-definite Gram matrix $X_{kt}$ constructed from the vectors $\{|v_m\rangle,\langle v_1|v_m\rangle|v_1\rangle,\langle v_2|v_m\rangle|v_2\rangle,\dots\}$, where $\Pi_i'=|v_i\rangle\langle v_i|$.
For this matrix, the experimental results indicate the constraint $|X_{kt}|^2\leq\epsilon_{mk}\epsilon_{kt}\epsilon_{mt}$ for all $k,t\geq1,k\neq t$ where $\epsilon_{ij}$ are initially set to 1 for $(i,j)\notin E$. We further assume a system dimension of $d=4$ for Peres-24, obtaining upper and lower bounds for the sum of $P_i$ within each context. A non-zero lower bound for any context indicates the successful verification of completeness. Additionally, we apply the constraint $\sum_{k\in V}X_{kk}-\sum_{k,t\in E}X_{kk}\epsilon_{kt}\geq \mathcal W_{\rm SDP}$ and apply a searching algorithm for minimal $\mathcal W_{\rm SDP}$ as the verification threshold. Furthermore, maximizing $\epsilon_{ij}$ for $(i,j)\notin E$ and iteratively updating its value in $|X_{kt}|^2$ tightens the constraints and reduces $\cal W_{\rm SDP}$. 
See the Supplemental Material \cite{sm} for further details. 

Finally, the robust certification succeeds if $\mathcal W_{\rm worst} > \mathcal W_{\rm SDP}$~\cite{XuZhenpeng}.
That is, when we can certify that the set of measurements is the target SI-C set up to the precision in the scale of square root of the experimental imperfection.


{\it Experiment and results.} The experiment setup is illustrated in Fig.~\ref{fig:setup}. A heralded single photon is generated with the wavelength at 808 nm \cite{fedrizzi2007wavelength}, see Fig.~\ref{fig:setup} (a). Its spatial modes and polarization degrees of freedom are used to encode two qubits. Specifically, the basis of Peres' set of 24 ququart observables are encoded as $|1\rangle=|LH\rangle,|2\rangle=|LV\rangle,|3\rangle=|RH\rangle,|4\rangle=|RV\rangle$, where $|L\rangle$ and $|R\rangle$ denote left and right spatial mode, respectively, and $|H\rangle$ and $|V\rangle$ denote horizontal and vertical polarization, respectively.


\begin{figure*}[thb]
\includegraphics[width=0.99\linewidth]{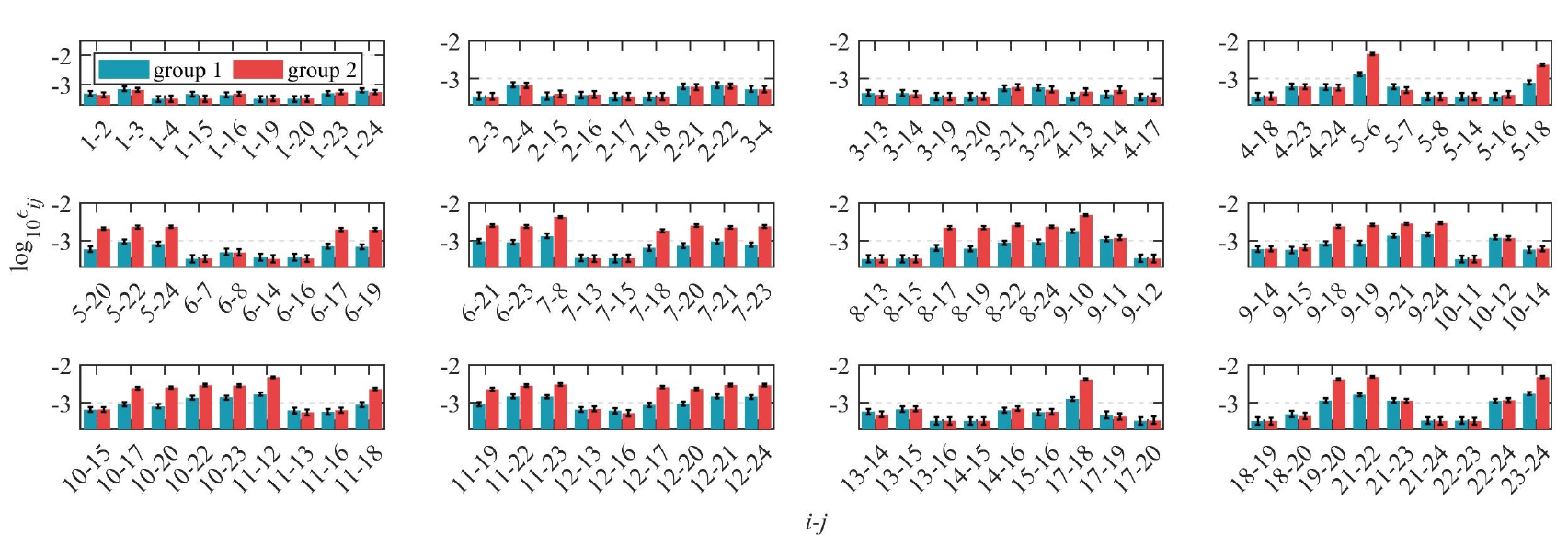}
\caption{Experiment outcomes of $\epsilon_{ij}$ obtained from projectors in group 1 and group 2, shown by green and red bars, respectively. Experimental error bars are estimated as the standard deviation based on assumption of photons' Poisson statistics.}
\label{fig:Result1}
\end{figure*}


The maximally mixed state is prepared in P1. Spatial modes are firstly prepared by using a beam displacer (BD) that splits the polarized components into different spatial modes with the help of a half-wave plate (HWP). Before the BD, a quartz crystal (QC) with a long thickness enough to reduce the coherence between polarized components is inserted. After the BD, the polarization in each mode is then adjusted by HWP followed by a QC to reduce the polarization coherence.

In M1, the maximally mixed state is measured in the polarizations and spatial modes according to the angles of the related motorized half wave plates (MHWP). From the measurement results, we obtain the projection basis $|v_i\rangle$ corresponding to the projector $\Pi_i'=|v_i\rangle\langle v_i|$ and its projection probability $P_i$. The system then automatically reprepares $|v_i\rangle$ using the MHWP in P2 with the same method in P1, followed by the measurement in M2 to obtain $\epsilon_{ij}$. Both probabilities of $P_i$ and $\epsilon_{ij}$ are obtained by analyzing the count from a given basis versus the sum of counts over an orthogonal complete set that includes that basis \cite{shalm2015strong,christensen2013detection}.

When analyzing the impact of experimental errors on the certification, a manual angular error is introduced into the final HWP in M2, and the total angular error is denoted as $\delta\theta$.

According to Peres-24, two groups of uncharacterized projectors (referred to as group 1 and 2) are performed on the maximally mixed state, respectively, where angular error is manually added to group 2.

\begin{figure}
\includegraphics[width=0.95\linewidth]{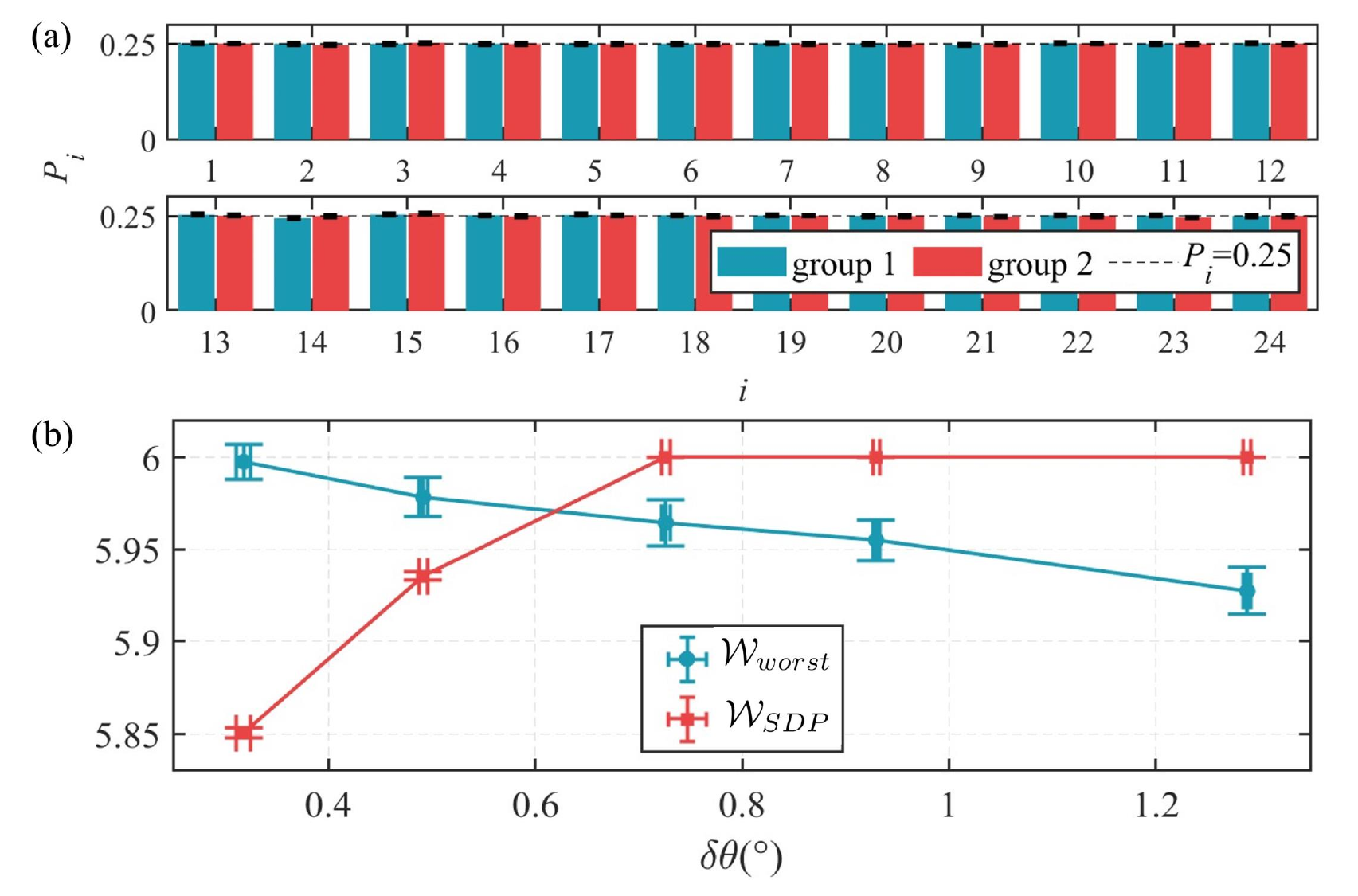}
\caption{Experimental results of $P_i$ obtained from group 1 and group 2 is shown by green and red bars in (a), respectively. In (b), we demonstrate $\mathcal W_{\rm worst}$ and $\mathcal W_{\rm SDP}$ with green circles and red squares, where the first and second column correspond to group 1 and 2, and other three columns correspond to three additional groups of projectors. Experimental error bars are estimated as the standard deviation based on assumption of photons' Poisson statistics. Horizontal error bars are on the order of $0.001$, whose gaps are not so obvious in the figure.}
\label{fig:Result2}
\end{figure}

The values of $\epsilon_{ij}$, the probability of projecting $|v_i\rangle$ onto orthogonal basis $|v_j\rangle$, are expected to be 0 theoretically for all $(i,j)\in E$, whose mean values obtained from group 1 and 2 equal $0.00071\pm0.00011$ and $0.00136\pm0.00014$, respectively, as shown in Fig.~\ref{fig:Result1}. The mean value of $\epsilon_{ij}$ in group 2 is higher than that in group 1 since the manual error is added to group 2. For example, $\epsilon_{5-6}=0.00131\pm0.00014$ and $\epsilon_{9-19}=0.00085\pm0.00012$ in group 1, while $\epsilon_{5-6}=0.00443\pm0.00027$ and $\epsilon_{9-19}=0.00256\pm0.00021$ in group 2. Additionally, experimental outcomes $P_i$ are shown in Fig.~\ref{fig:Result2} (a), which are expected to equal $0.25$ in theory and achieved with the mean value of $0.2504\pm0.0021$ and $0.2496\pm0.0020$ in group 1 and 2, respectively. Our method demands a low level of $\epsilon_{ij}$ for which the noise introduced by experimental components are eliminated, detailed in the Supplemental Material \cite{sm}. The worst witness and threshold witness for group 1 and group 2 are evaluated as $\mathcal W_{\rm worst}=5.997\pm0.010,\ \mathcal W_{\rm SDP}=5.8504\pm0.0026$ and $\mathcal W_{\rm worst}=5.978\pm0.011,\ \mathcal W_{\rm SDP}=5.9353\pm0.0022$, respectively, revealing that $\mathcal W_{\rm worst}>\mathcal W_{\rm SDP}$ for both groups, which indicates that the uncharacterized projectors of group 1 and group 2 are successfully certified as Peres-24 in our experimental system.

Furthermore, we perform three additional groups of uncharacterized sequential projection measurements with increasing manual angular offsets. 
In addition, an optimization algorithm is performed to evaluate the error level by attributing all error to $\delta\theta$, and details could be found in the Supplemental Material \cite{sm}. $\delta\theta$ obtained from group 1 and 2 are $0.318^\circ\pm0.007^\circ$ and $0.492^\circ\pm0.004^\circ$. 
The worst witness $\mathcal W_{\rm worst}$ decreases as the $\delta\theta$ increases, while the threshold witness $\mathcal W_{\rm SDP}$ performs diametrically, as shown in Fig.~\ref{fig:Result2} (b). Specifically, as introduced above, projectors in group 1 and group 2 are successfully certified to be the Peres-24 set with $\mathcal W_{\rm worst}>\mathcal W_{\rm SDP}$ as shown in the first and second columns of Fig.~\ref{fig:Result2} (b). 
For the last three columns, we observe $\mathcal W_{\rm worst}<\mathcal W_{\rm SDP}$. And $\mathcal W_{\rm SDP}$ reaches $\mathcal W_{\rm opt}$, indicating that certification is impossible for these groups with such an error level. These results demonstrate the scheme's sensitivity, as even minor errors cause a significant increase in $\mathcal W_{\rm SDP}$, leading to certification failure. Conversely, successful certification implies that the imperfections of the uncharacterized projectors are below a strict threshold, providing evidence that they closely approximate the ideal SI-C set.


{\it Discussion.} We have presented a new theoretical method for certifying specific sets of quantum measurements using only maximally mixed states. We have demonstrated its feasibility by successfully experimentally certifying in a photonic setup two fundamental sets of measurements in quantum theory: Peres-24 \cite{Peres24} in dimension four, and---as detailed in the Supplemental Material---Yu-Oh-13 \cite{YO13} in dimension three. The method is equally applicable to many other SI-C sets in any finite dimension $d \ge 3$ (e.g., \cite{BBC-21,CEG-18,PM-Peres,PM-Mermin,PM-Cabello,XuZhenpeng}). The observables certified are themselves significant, as SI-C sets have applications in quantum key distribution \cite{QKDConGupta,QKDConSingh}, quantum metrology \cite{jae2024contextual}, and quantum computation \cite{howard2014contextuality}.

In contrast to conventional certification methods based on nonlocality and state-dependent contextuality, which rely on the preparation of specific pure states to attain extremal points of the quantum correlations, our method requires only preparing a maximally mixed state. This significantly mitigates the negative effect of imperfections in state preparation as the maximally mixed state can be prepared in infinitely many ways. Interestingly, the method is particularly advantageous in high dimensions because preparing maximally mixed states becomes significantly easier than generating specific pure states as the dimension grows.

In addition, the method can be extended to other full-rank mixed states. For example, we can use thermal states, which are commonly encountered in various physical systems. This flexibility broadens the applicability of the method across diverse experimental platforms \cite{wei2010testing,ConExpKirchmair,zhang2013state}.


J.-H. Liang and Z.-Y. Hao contribute equally to this work. We thank Armin Tavakoli for useful discussions. This work was supported by National Key Research and Development Program of China (Grant No. 2025YFE0217700), Quantum Science and Technology-National Science and Technology Major Project (Grants No.\ 2021ZD0301200 and No.\ 2021ZD0301400), the National Natural Science Foundation of China (Grants Nos.\ 92365205, 12374336, W2411001, 123B2067, and 62475249), the Anhui Provincial Natural Science Foundation (Grant No.\ 2508085J002), USTC Major Frontier Project (LS2030000002), and USTC Research Funds of the Double First-Class Initiative (Grant No.\ YD2030002024). Z.-P.\ X.\ acknowledges support from National Natural Science Foundation of China (No.\ 12305007), Anhui Provincial Natural Science Foundation (Nos.\ 2308085QA29 and 2508085Y003), Anhui Province Science and Technology Innovation Project (No.\
202423r06050004). A.\ C.\ was supported by EU-funded project \href{10.3030/101070558}{FoQaCiA}. 


\bibliography{abbreviated}

\end{document}


\title{Experimental Certification of Quantum Measurements with Maximally Mixed States. Supplemental Material}


\author{Jia-He Liang}

\author{Ze-Yan Hao}

\author{Jia-Kun Li}

\affiliation{Laboratory of Quantum Information, University of Science and Technology of China, Hefei 230026, China}
\affiliation{Anhui Province Key Laboratory of Quantum Network, University of Science and Technology of China, Hefei 230026, China}
\affiliation{CAS Center for Excellence in Quantum Information and Quantum Physics, University of Science and Technology of China, Hefei 230026, China}

\author{Kai Sun}

\email{ksun678@ustc.edu.cn}
\affiliation{Laboratory of Quantum Information, University of Science and Technology of China, Hefei 230026, China}
\affiliation{Anhui Province Key Laboratory of Quantum Network, University of Science and Technology of China, Hefei 230026, China}
\affiliation{CAS Center for Excellence in Quantum Information and Quantum Physics, University of Science and Technology of China, Hefei 230026, China}
\affiliation{Hefei National Laboratory, University of Science and Technology of China, Hefei 230088, China}

\author{Zhen-Peng Xu}
\email{zhen-peng.xu@ahu.edu.cn}

\affiliation{School of Physics and Optoelectronics Engineering, Anhui University, 230601 Hefei, China}

\author{Jin-Shi Xu}
\email{jsxu@ustc.edu.cn}

\author{Chuan-Feng Li}
\email{cfli@ustc.edu.cn}

\author{Guang-Can Guo}

\affiliation{Laboratory of Quantum Information, University of Science and Technology of China, Hefei 230026, China}
\affiliation{Anhui Province Key Laboratory of Quantum Network, University of Science and Technology of China, Hefei 230026, China}
\affiliation{CAS Center for Excellence in Quantum Information and Quantum Physics, University of Science and Technology of China, Hefei 230026, China}
\affiliation{Hefei National Laboratory, University of Science and Technology of China, Hefei 230088, China}

\author{Ad\'{a}n Cabello}
\email{adan@us.es}
\affiliation{Departamento de Física Aplicada II, Universidad de Sevilla, E-41012 Sevilla, Spain}
\affiliation{Instituto Carlos I de Física Teórica y Computacional, Universidad de Sevilla, E-41012 Sevilla, Spain}

\date{\today}%

\begin{abstract}
	Section~\ref{sec:I} details the two sets of measurements tested, Peres-24 and YO-13. Section~\ref{sec:II} provides additional experimental details, while Sec.~\ref{sec:III} describes the noise elimination method. Section~\ref{sec:IV} details the worst-case witness, Sec.~\ref{sec:V} the semi-definite program, Sec.~\ref{Sec:VI} the results of the experiment for YO-13, and Sec.~\ref{sec:VII} the calculation of $\delta\theta$.
\end{abstract}


\maketitle


\section{The two sets of measurements tested}
\label{sec:I}


\subsection{Peres-24}


Peres-24 is a set of 24 observables represented by rank-one projectors in dimension $d=4$ introduced in Ref.~\cite{Peres24}. The elements of Peres-24 and their respective weights in the SI-C witness $\mathcal W$ [see Eq.~(1) in the main text] are shown in Table~\ref{tab:Peres24}. The 24 vectors of Peres-24 form 24 orthogonal bases. Each vector is in four orthogonal bases [see Fig.~1 (b) in the main text]. As proven in Ref.~\cite{XuZhenpeng}, Peres-24 is unique up to unitary transformations \cite{XuZhenpeng}.

When calculating the experimental witness, we set weights of each edge in Eq.~(1) in the main text to $w_{ij}=1$. This satisfies $w_{ij}\geq {\rm max}\{w_i,w_j\}$. With these weights, we can theoretically obtain the ideal witness $\mathcal W_{\rm opt}=6$.


\begin{table*}[h]
	\renewcommand\arraystretch{1.2}
	\begin{tabular}{ccccccccccccccccccccccccc}
		\hline\hline
		~ & $v_1$ & $v_2$ & $v_3$ & $v_4$ & $v_5$ & $v_6$ & $v_7$ & $v_8$ & $v_9$ & $v_{10}$ & $v_{11}$ & $v_{12}$ & $v_{13}$ & $v_{14}$ & $v_{15}$ & $v_{16}$ & $v_{17}$ & $v_{18}$ & $v_{19}$ & $v_{20}$ & $v_{21}$ & $v_{22}$ & $v_{23}$ & $v_{24}$\\
		\hline
		$v_{i1}$ & $1$ & $0$ & $0$ & $0$ & $1$ & $1$ & $1$ & $1$ & $1$ & $1$ & $1$ & $\bar1$ & $1$ & $1$ & $0$ & $0$ & $1$ & $1$ & $0$ & $0$ & $1$ & $1$ & $0$ & $0$\\
		$v_{i2}$ & $0$ & $1$ & $0$ & $0$ & $1$& $1$& $\bar1$& $\bar1$ & $1$& $1$& $\bar1$& $1$& $1$& $\bar1$& $0$& $0$& $0$& $0$& $1$& $1$ &$0$& $0$& $1$& $1$\\
		$v_{i3}$ & $0$& $0$& $1$& $0$& $1$& $\bar1$& $1$& $\bar1$ & $1$& $\bar1$& $1$& $1$& $0$& $0$& $1$& $1$& $1$& $\bar1$& $0$& $0$& $0$& $0$& $1$& $\bar1$\\
		$v_{i4}$ & $0$& $0$& $0$& $1$& $1$& $\bar1$& $\bar1$& $\bar1$& $1$& $1$& $1$& $1$& $0$& $0$& $1$& $\bar1$& $0$& $0$& $1$& $\bar1$ & $1$& $\bar1$ & $0$& $0$\\
		\hline
		$w_i$ & $1$& $1$& $1$& $1$& $1$& $1$& $1$& $1$& $1$& $1$& $1$& $1$& $1$& $1$& $1$& $1$& $1$& $1$& $1$& $1$& $1$& $1$& $1$& $1$\\
		\hline\hline
	\end{tabular}
	\caption{Peres-24 has 24 observables represented by rank-one projectors $|v_i\rangle\langle v_i|$. Each column $v_i$ contains the (non-normalized) components of $|v_i\rangle$. For simplicity, $\bar1=-1$. The last row, $w_i$, indicates the weight of $v_i$ in the SI-C witness in Eq.~(1) in the main text.}
	\label{tab:Peres24}
\end{table*}


\subsection{YO-13}

\begin{figure}[h]
	\centering
	\includegraphics[width=0.45\linewidth]{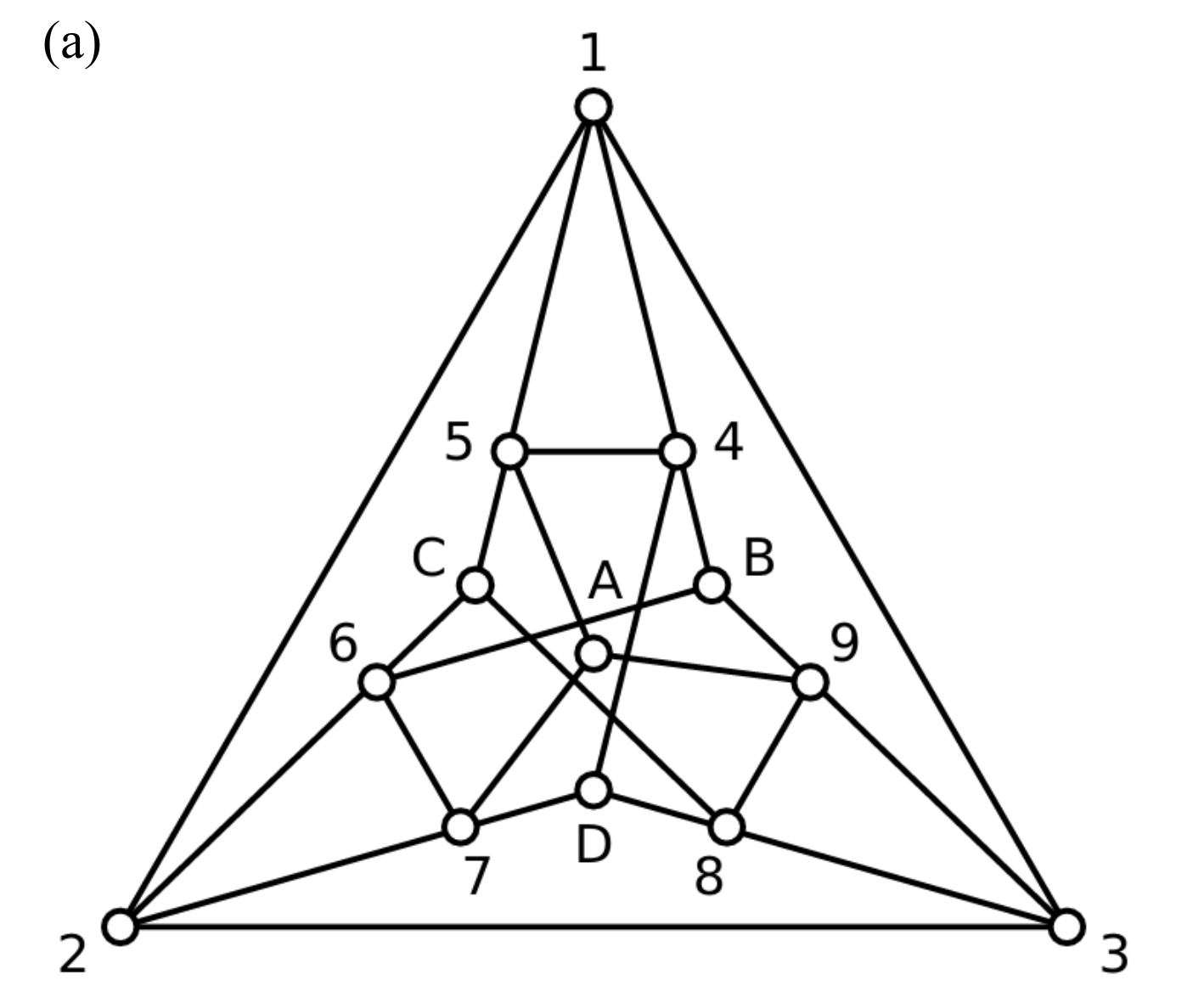}
	\includegraphics[width=0.45\linewidth]{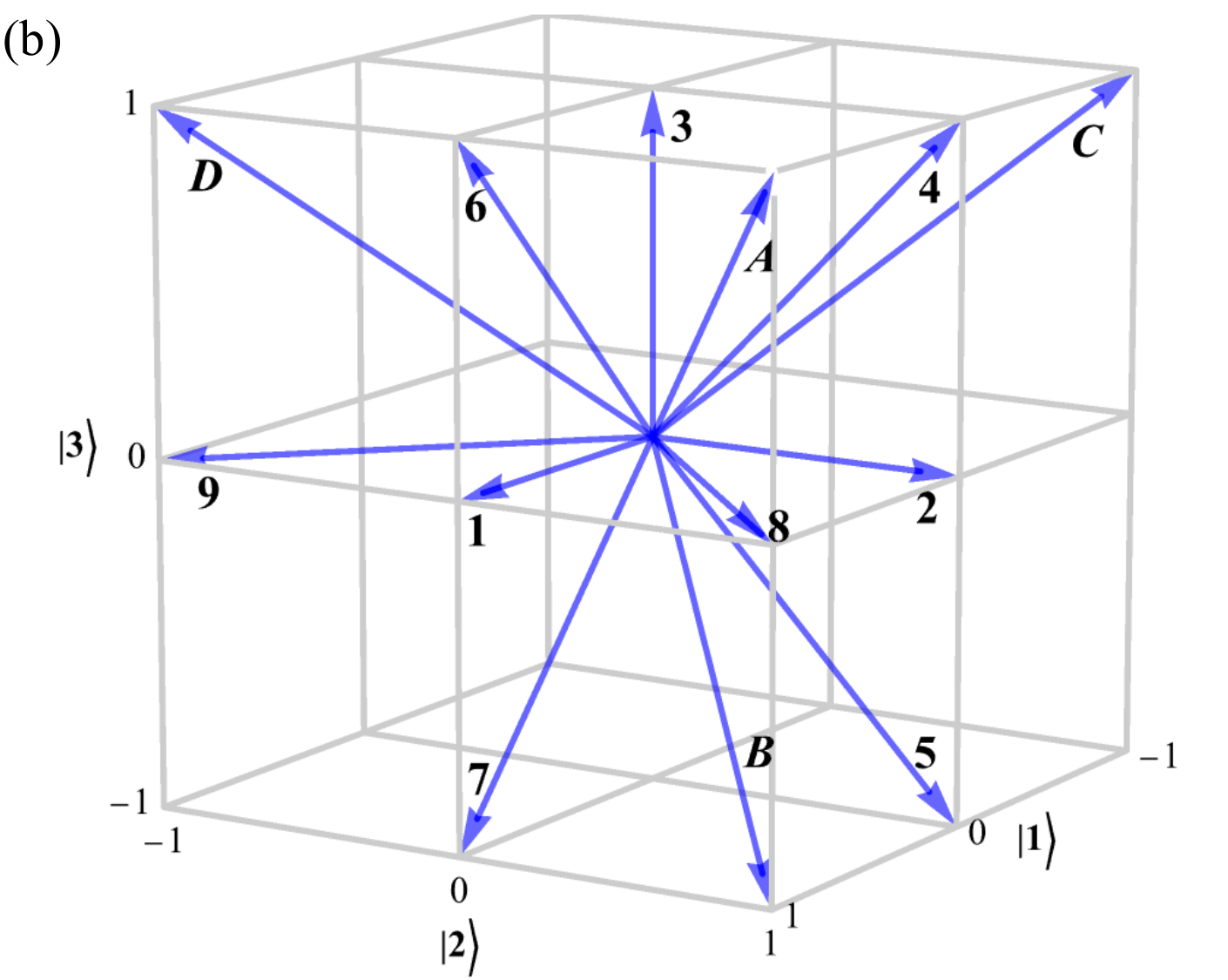}
	\caption{ (a) Orthogonality graph for the YO-13 set \cite{YO13} given in Table~\ref{tab:YO13}. Vertices represent measurements. Vectors in the same straight line are mutually orthogonal. (b)~Representation of the (non-normalized) vectors of YO-13 in the Euclidean space. }
	\label{fig:yo13}
\end{figure}

YO-13 is a set of 13 observables represented by rank-one projectors in dimension $d=3$ introduced in Ref.~\cite{YO13}. The elements of YO-13 and their respective weights in the SI-C witness $\mathcal W$ [see Eq.~(1) in the main text] are shown in Table~\ref{tab:YO13}. Its orthogonality graph is shown in Fig.~\ref{fig:yo13} (a).


\begin{table}[h]
	\renewcommand\arraystretch{1.2}
	\begin{tabular}{cccccccccccccc}
		\hline\hline
		~ & $v_1$ & $v_2$ & $v_3$ & $v_4$ & $v_5$ & $v_6$ & $v_7$ & $v_8$ & $v_9$ & $v_{A}$ & $v_{B}$ & $v_{C}$ & $v_{D}$\\
		\hline
		$v_{i1}$ & $1$ & $0$ & $0$ & $0$ & $0$ & $1$ & $1$ & $1$ & $1$ & $1$ & $1$ & $\bar1$ & $1$\\
		$v_{i2}$ & $0$ & $1$ & $0$ & $1$ & $1$ & $0$ & $0$ & $1$ & $\bar1$ & $1$ & $1$ & $1$ & $\bar1$\\
		$v_{i3}$ & $0$ & $0$ & $1$ & $1$ & $\bar1$ & $1$ & $\bar1$ & $0$ & $0$ & $1$ & $\bar1$ & $1$ & $1$\\
		\hline
		$w_{i}$ & $3$ & $3$ & $3$ & $3$ & $3$ & $3$ & $3$ & $3$ & $3$ & $2$ & $2$ & $2$ & $2$\\
		\hline\hline
	\end{tabular}
	\caption{YO-13 has 13 observables represented by rank-one projectors $|v_i\rangle\langle v_i|$. Each column $v_i$ contains the (non-normalized) components of $|v_i\rangle$. For simplicity, $\bar1=-1$. The last row, $w_i$, indicates the weight of $v_i$ in the SI-C witness in Eq.~(1) in the main text.}
	\label{tab:YO13}
\end{table}


The weights of the edges in Eq.~(1) in the main text are set to $w_{ij}=3$. This satisfies $w_{ij}\geq {\rm max}\{w_i,w_j\}$. With these assumptions we can theoretically obtain $\mathcal W_{\rm opt}=35/3$ in $d=3$.



YO-13 has only four complete orthogonal bases: $\{v_1,v_2,v_3\}$, $\{v_1,v_4,v_5\}$, $\{v_2,v_6,v_7\}$, and $\{v_3,v_8,v_9\}$. Vectors $v_A$, $v_B$, $v_C$, and $v_D$ are not in any of these sets. To ensure that every projector is experimentally measured as part of a complete basis, we add four unnormalized vectors, $v_E=(2,-1,-1)$, $v_F=(-1,2,1)$, $v_G=(1,-1,2)$, and $v_H=(1,2,1)$. This allows us to define four new complete orthogonal bases: $\{v_5,v_A,v_E\}$, $\{v_6,v_B,v_F\}$, $\{v_8,v_C,v_G\}$, and $\{v_7,v_D,v_H\}$. Vectors $v_E$, $v_F$, $v_G$, and $v_H$ are only considered when implementing experimental measurements, but they are not considered in the SDP.

As proven in Ref.~\cite{XuZhenpeng}, YO-13 is unique up to unitary transformations if 
\begin{subequations}
	\begin{align}
		P_1+P_2+P_3&=1,\\
		P_4+P_5+P_6&=1,\\
		P_7+P_8+P_9&=1,
	\end{align}
	\label{eq:yo13id}
\end{subequations} 
which is satisfied in $d=3$.


\section{Experimental details}
\label{sec:II}


A heralded single photon source is set for the experiments, shown in Fig.~1 (b) in main text. There is a 404 nm continuous wave laser pumping a 2 cm-length periodically poled Potassium Titanyl Phosphate crystal, generating horizontal and vertical polarized 808 nm photons with type-II quasi-phase matching. Two photons are separated by a polarization beam splitter (PBS) in front of the optical system. The vertically polarized photon---called trigger photon---is reflected and coupled into a fiber connected with a single photon avalanche diode (SPAD). The horizontally polarized photon---called heralded photon---passes through the PBS and enters the optical system. The heralded photon that exits from block M2 is coupled into a fiber connected with another SPAD. Generally, if both SPADs detect the photons generated from single photon source, then the signal of heralded photon is a few nanoseconds delayed from trigger photon since heralded photon goes a longer optical path length determined by the optical system. Based on such feature, the coincidence counter counts the coincident events with such delay and obtains the number of such coincident events in a period of time (10 s in our experiments).

When implementing sequential measurements of YO-13, states $|1\rangle$, $|2\rangle$, and $|3\rangle$ are encoded as $|LH\rangle$, $|LV\rangle$, and $|RV\rangle$, respectively.

Angles of the half-wave plates (HWPs) are automatically controlled by motors and determined by the states to prepare and project. The angle of the first HWP in block P2 is denoted as $\theta_1$ and the angles of the HWP in the left and right paths are denoted as $\theta_2$ and $\theta_3$, respectively. According to the experimental setup and the Jones matrix of the HWP, 
\begin{equation}
	J(\theta)=
	\begin{pmatrix}
		\cos2\theta & \sin2\theta\\
		\sin2\theta & -\cos2\theta
	\end{pmatrix},
	\label{eq:jones}
\end{equation}
the polarization of the heralded photon is rotated by the first HWP into $(\cos2\theta_1,\sin2\theta_1)^T$ and these two components are split into two paths, which encode the spatial mode qubit. The state obtained behind the beam displacer (BD) of P2 block is $(\cos2\theta_1,0,0,\sin2\theta_1)^T$. The polarization qubit is encoded by the following HWP and the four component of the state finally prepared are $v_{i1}=\cos2\theta_1\cos2\theta_2$, $v_{i2}=\cos2\theta_1\sin2\theta_2$, $v_{i3}=\sin2\theta_1\sin2\theta_3$, $v_{i4}=-\sin2\theta_1\cos2\theta_3$. In M2, we denote the angle of the last HWP in M2 as $\theta_4$, and $\theta_5$, $\theta_6$ are denoted as the angles for the HWPs in the left and right spatial modes in front of the two PBSs, respectively. Similarly, the components of the state to project in M2 are $v_{j1}=\sin2\theta_4\cos2\theta_5$, $v_{j2}=\sin2\theta_4\sin2\theta_5$, $v_{j3}=\cos2\theta_4\cos2\theta_6$, $v_{j4}=\cos2\theta_4\sin2\theta_6$. The HWP between the PBS and the BD in the left path is fixed at $45^\circ$ to rotate the horizontal polarized photons that go through the PBS into vertical, to combine with the other path in BD. The setups in the M1 block are similar to those in the M2 block.


\section{Noise elimination}
\label{sec:III}


Certifying the set of measurements requires a strict threshold for the experimental imperfections \cite{XuZhenpeng}. Therefore, we have made efforts to minimize the experimental imperfections. Taking group 1 as an example, the minimum $\epsilon_{ij}$ falls below $1/3000$. However, because of the quantum noise introduced by the BD and the PBS, the experimental outcomes do not directly represent the actual error of the projection measurements. We therefore adopt a noise model to estimate the noise level, which is subsequently eliminated to obtain the actual outcomes of the projection measurements. Details of the noise elimination procedure are provided below.

In the remainder of this Supplemental Material, we denote the ideal projector as $\Pi_i=|v_i\rangle\langle v_i|$ and those noiseless projections actually performed in the system as $\Pi'_i=|V_i\rangle\langle V_i|$. Similarly, the state reprepared in block P2 is denoted as $|V_i\rangle$. Due to the experimental imperfections, $|V_i\rangle$ are slightly different from the ideal state $|v_i\rangle$, whose unnormalized form is denoted as
\begin{equation}
	|\tilde V_i\rangle=|v_i\rangle+|\Delta i\rangle
\end{equation}
and $|V_i\rangle=|\tilde V_i\rangle/\sqrt{\langle \tilde V_i|\tilde V_i\rangle}$. $\{|\Delta i\rangle\}$ denotes the minor error from the ideal states.

Two noise channels are considered in our experiment setup. Due to the geometry of the PBS cube and the angular dependence of the beam-splitting interface coatings, a fraction of vertically polarized component is leaked in transmission, resulting a bit-flip channel in both the polarization and path qubits \cite{PBSangular}. In addition, there is a Mach–Zehnder interferometer in P2 and M2, where photon is split into two spatial modes in P2 and combined in M2. Minor optical length differences and spatial separation may reduce the coherence of the two spatial modes introducing a phase-flip channel \cite{MZdecoherence}. The Kraus operators are
\begin{subequations}
	\begin{align}
		E^{Pa}_0 &=\sqrt{1-p_{Pa}} \ I\otimes I,\;\; 
		&E^{Pa}_1 =\sqrt{p_{Pa}} \ \sigma_z\otimes I,\\
		E^{Ba}_0 &=\sqrt{1-p_{Ba}} \ I\otimes I,\;\;
		&E^{Ba}_1 =\sqrt{p_{Ba}} \ \sigma_x\otimes I,\\
		E^{Bb}_0 &=\sqrt{1-p_{Bb}} \ I\otimes I,\;\; &
		E^{Bb}_1 =\sqrt{p_{Bb}} \ I\otimes \sigma_x,
\end{align}
\label{eq:krauss}
\end{subequations}
where $B$ and $P$ denotes bit-flip and phase-flip, respectively. $\sigma_x$ and $\sigma_z$ are Pauli operators and $I$ is the identity operator. $a$ and $b$ denote the channels applied on the spatial mode qubit and the polarization qubit, respectively, and $p_{Ba}$, $p_{Bb}$, and $p_{Pa}$ are the probabilities of the corresponding noise channel. The state obtained from applying these noise channels on $|V_i\rangle$ is 
\begin{equation}
	\rho_i=\sum_{m,n,v\in\{0,1\}}E_m^{Ba}E_n^{Bb}E_v^{Pa} |V_i\rangle\langle V_i| E_v^{\dagger Pa}E_n^{\dagger Bb}E_m^{\dagger Ba}.
\end{equation}
The noisy probabilities of the projection $i-j$ are $\langle V_j|\rho_i|V_j\rangle$. Now we implement an optimization program of
\begin{equation}
	\begin{aligned}
		\min_{\{|\Delta i\rangle\},p_{Pb},p_{Ba},p_{Bb}} \ &\sum_{i,j\in E}[\langle V_j|\rho_i|V_j\rangle-\epsilon_{ij}]^2\\
		s.t. \ &0\leq p_{Ba},p_{Bb}\leq1,\\
		& 0\leq p_{Pa}\leq \frac{1}{2},
	\end{aligned}
\end{equation}
where $\epsilon_{ij}$ are the noisy experiment outcomes. When the optimum is achieved, we can obtain the $|V_i\rangle$ and the parameters of the noise channels that better match the experiment. Therefore, we obtain the noiseless error of pure states $\epsilon'_{ij}=|\langle V_j|V_i\rangle|^2$. This method is used only for obtaining noiseless $\epsilon'_{ij}$ rather $|V_i\rangle$ to be certified. It still requires SDP and calculating $\cal W_{\rm worst}$ to determine the certification. Take group 1 as example, SDP is performed on $\epsilon'_{ij}$ and finally obtain $\mathcal W_{SDP}=5.8504\pm0.0026$, while the worst-case witness is evaluated as $\mathcal W_{worst}=5.997\pm0.010$. This result corresponds to a successful certification since $\mathcal W_{worst}>\mathcal W_{SDP}$.


\section{Details on the worst-case witness $\mathcal W_{\rm worst}$}
\label{sec:IV}


There are two factors that make $\mathcal W_{\rm opt}$ experimentally unachievable: the measurement imperfection and the initial state imperfection. 

We first discuss the measurement imperfection.
According to Eq.~(1) in the main text and to the experimental method for obtaining $P_{ij}$ in Ref.~\cite{XiaoYa}, we set a witness operator 
\begin{equation}
	\hat{\mathcal W}=\sum_{i\in V}w_i\Pi_i'-\sum_{i,j\in E}w_{ij}\Pi_i'\Pi_j'\Pi_i', \label{eq:witness}
\end{equation}
such that the witness in Eq.~(1) in the main text is obtained as 
\begin{equation}
\mathcal W(\rho)={\rm tr}(\rho\hat{\mathcal W}),
\end{equation}
where $\rho$ denotes the input state. In the ideal case of $\Pi_i'=\Pi_i$, the second item in the right-hand side of \eqref{eq:witness} equals zero, obtaining $\mathcal W(\rho)=\mathcal W_{\rm opt}$ for all input states $\rho$. 

Now we diagonalize the experimental witness $\hat{\mathcal W}$ as
\begin{equation}
	\hat{\mathcal W}=\sum_k^d \lambda_k|k\rangle\langle k|,
\end{equation}
where $\{|k\rangle\}_{k=1}^d$ are eigenvectors in the $d$-dimension Hilbert space and $\lambda_k$ denotes the eigenvalues of each basis. Then the witness obtained from a maximally mixed state $\rho_{\rm mm}$ equals 
\begin{equation}
\mathcal W(\rho_{\rm mm})={\rm tr}(\rho_{\rm mm}\sum_k^d \lambda_k|k\rangle\langle k|)=\sum_{k}^d\lambda_k\langle k|\rho_{\rm mm}|k\rangle.
\end{equation}
Taking into account that, for a maximally mixed state, any probability of obtaining a pure state is $1/d$, 
the witness $\mathcal W(\rho_{\rm mm})$ obtained with measurement imperfection is
\begin{equation}
	\mathcal W_{\rm exp}=\sum_{k}^d\lambda_k/d. \label{eq:exp}
\end{equation}
Here we assume that $\lambda_{k=d}$ is the minimum value in all $\{\lambda_k\}$.

Now we discuss the imperfection in the initial state. Let us consider the implementation of $\hat{\mathcal W}$ on an arbitrary state $\rho$, whose witness equals
\begin{equation}
	\mathcal W(\rho)={\rm tr}(\rho\hat{\mathcal W})=\sum_{k}^d\lambda_k\langle k|\rho|k\rangle,
\end{equation}
from which the minimum $\mathcal W(\rho)=\lambda_{k=d}$ is obtained only when $\rho=|d\rangle\langle d|$. Equivalently, the witness that is possibly obtained from any state from the experiment is not less than $\lambda_d$, which is considered the worst-case witness $\mathcal W_{\rm worst}$. However, eigenvalues $\{\lambda_k\}$ are not experimentally obtainable in this work since we only implement $\hat{\mathcal W}$ on the maximally mixed state. To address this issue, we evaluate a lower bound of $\lambda_d$ that satisfies constraint of (\ref{eq:exp}). It is assumed that all $\lambda_k=\mathcal W_{\rm opt}$ for $k=1,\ldots,d-1$, and all experimental error of $\mathcal W_{\rm exp}$ are introduced from $\lambda_d$. Subsequently, the worst case of the minimum witness $\lambda_d$, denoted $\mathcal W_{worst}$, is finally evaluated as 
\begin{equation}
	\mathcal W_{\rm worst}=\mathcal W_{\rm opt}-d(\mathcal W_{\rm opt}-\mathcal W_{\rm exp}),
\end{equation}
which equals to Eq.~(2) in the main text by expanding $\mathcal W_{\rm exp}$.

According to the derivation above, the minimum witness of testing over all states is $\lambda_{k=d}$ rather $\mathcal W_{\rm worst}$ and $\lambda_{k=d}\geq\mathcal W_{\rm worst}$. Based on this characteristic, there is a possibility of $\lambda_{k=d}>\mathcal W_{\rm SDP}>\mathcal W_{\rm worst}$, which shows a failure of certification in our scheme but is theoretically able to be certified. Therefore, the positive result of $\mathcal W_{\rm worst}>\mathcal W_{\rm SDP}$ in this scheme is sufficient (but not necessary) for a certification.


\section{Details on the semi-definite program and $\mathcal W_{\rm SDP}$}
\label{sec:V}


The orthogonality and completeness of $\{\Pi_i'\}$ are analyzed using $\epsilon'_{ij}$ in a semi-definite program (SDP), giving a certification threshold $\mathcal W_{\rm SDP}$.

Notice that, in the second line of Eq.~(2) in the main text, the values of $\epsilon'_{ij}$ are determined by only projectors $\{\Pi_i'\}$ and are independent on the initial states, indicating that $\epsilon'_{ij}$ is the same for every initial states. Therefore, we assume that $\epsilon'_{ij}$ for the worst result are the same as those obtained from the maximally mixed state.

We first verify the completeness of the projectors in each complete and orthogonal context $c$. In this verification, we have assumed the dimension $d$ according to the physical feature of the experimental system (two spatial modes and two polarization components), so that we can use method demonstrated as follows. For a complete context, The sum of projectors is denoted as $\Sigma=\sum_{k}^d |V_k\rangle\langle V_k|$, where the rank of $\Sigma$ equals $d$ indicating the minimum eigenvalue $\lambda_{min}(\Sigma)>0$. Here we introduce Gram matrix $G$ with its component $G_{ij}=\langle V_i|V_j\rangle$ whose eigenvalues are the same as $\Sigma$. We can get the maximum evaluation of eigenvalues $g_{max}$ and the minimum evaluation of eigenvalues $g_{min}$ as
\begin{equation}
	\begin{aligned}
		g_{max}&=\max \lambda_{max}(G)\\
		s.t. &\ |G_{ij}|^2\leq\epsilon'_{ij}, \ (i,j)\in E,\\
		& \ G_{ii}=1, \ 1\leq i\leq d,\\
	\end{aligned}
\end{equation}
\begin{equation}
	\begin{aligned}
		g_{min}&=\min \lambda_{min}(G)\\
		s.t. &\ |G_{ij}|^2\leq\epsilon'_{ij}, \ (i,j)\in E,\\
		& \ G_{ii}=1, \ 1\leq i\leq d.\\
	\end{aligned}
\end{equation}
The completeness of the projectors is verified if $g_{min}>0$ for all complete context. 

Next, we use SDP to verify that the probability of projecting basis corresponding to unconnected vertices is greater than zero. We use another positive semi-definite Gram matrix $X_{kt}$ constructed from the vectors $\{|V_m\rangle,\langle V_1|V_m\rangle|V_1\rangle,\langle V_2|V_m\rangle|V_2\rangle,\dots\}$, then
\begin{equation}
	\begin{aligned}
		\tau_{ij}:=&\min X_{ij}\\
		s.t.\quad&\sum_{k=1}^nw_kX_{kk}-\sum_{k,t\in E}w_{kt}X_{kk}\epsilon'_{kt}\geq\mathcal W_{\rm SDP},\\
		&X_{kk}=X_{0k}, \ 1\leq k\leq n,\\
		&X_{ij}\geq0, \ X_{00}=X_{0i}=1,\\
		&|X_{kt}|^2\leq\epsilon'_{ik}\epsilon'_{kt}\epsilon'_{tj}, \ 1\leq i,j\leq n, \ i\neq j,\\
		&g_{min}\leq\sum_{k\in c} X_{kk}\leq g_{max}, \ c\in C, \\
		&X\in\mathcal S_+^{1+n},
	\end{aligned}
	\label{tauij}
\end{equation}
where $\tau_{ij}$ denotes $|\langle V_i|V_j\rangle|^2$ of non-orthogonal basis $(i,j)\notin E$ and $C$ denotes the set of all complete and orthogonal context. Details of $\epsilon'_{ij}$ for $(i,j)\notin E$ is shown in the next paragraph. If any one of $\tau_{ij},(i,j)\notin E$ equals zero (less than $10^{-4}$ in practice), this verification of orthogonality fails. By adjusting the value of $\mathcal W_{\rm SDP}$, we can find a threshold above which the orthogonality is verified and fails below.

$\epsilon'_{ij}$ denotes the maximum constraint of $|\langle V_i|V_j\rangle|^2$ for all $(i,j)$. For those $(i,j)\in E$, the maximum constraint is determined by the experimental results. And for those $(i,j) \notin E$, the constraint of $\epsilon'_{ij}$ can be iteratively stringent by 
\begin{equation}
	\begin{aligned}
		\epsilon'_{ij}:=&\max X_{ij}\\
		s.t.\quad&\sum_{k=1}^nw_kX_{kk}-\sum_{k,t\in E}w_{kt}X_{kk}\epsilon'_{kt}\geq\mathcal W_{\rm SDP},\\
		&X_{kk}=X_{0k}, \ 1\leq k\leq n,\\
		&X_{ij}\geq0, \ X_{00}=X_{0i}=1,\\
		&|X_{kt}|^2\leq\epsilon'_{ik}\epsilon'_{kt}\epsilon'_{tj}, \ 1\leq i,j\leq n, \ i\neq j,\\
		&g_{min}\leq\sum_{k\in c} X_{kk}\leq g_{max}, \ c\in C, \\
		&X\in\mathcal S_+^{1+n}.
	\end{aligned}
	\label{epsilonprime}
\end{equation}
In the iteration, the start value of $\epsilon'_{ij}, \ (i,j) \notin E$ is set as $1$ and the SDP of \eqref{epsilonprime} is implemented on all $\epsilon'_{ij}$, obtaining the first values of $0\leq\epsilon'_{ij}\leq1$. Next, the SDP is performed based on the first values of $\epsilon_{ij}'$ and returns second values of $\epsilon_{ij}'$. According to the practical experience, $\epsilon_{ij}^{'}$ converge in the third values with precision of $0.001$. 

Specifically, YO-13 is unique up to unitary transformation when satisfying \eqref{eq:yo13id}. Therefore, when implementing SDP \eqref{tauij} and \eqref{epsilonprime} on experimental outcomes of YO-13, both $g_{max}$ and $g_{min}$ are manually set as $1$.

Without the constraint of system dimension, we can use another SDP method to verify the completeness.
\begin{equation}
	\begin{aligned}
		\nu:=&\min \sum_{k}^c X_{0k}\\
		s.t.\quad&\sum_{k=1}^nw_kX_{kk}-\sum_{k,t\in E}w_{kt}X_{kk}\epsilon'_{kt}\geq\mathcal W,\\
		&X_{kk}=X_{0k}, \ 1\leq k\leq n,\\
		&X_{00}=X_{0i}=1,\\
		&|X_{kt}|\leq\epsilon_{kt}, \ (k,t)\in E,\\
		&X\in\mathcal S_+^{1+n},
	\end{aligned}
\end{equation}
where $\nu$ denotes the probability sum of projecting to a complete context $c$. If $\nu=0$ for any of the $c$ is obtained from any one of $c$, this verification of completeness fails, otherwise succeed.


\section{Results of the experiment on YO-13}
\label{Sec:VI}


\begin{figure}[h]
	\centering
	\includegraphics[width=0.75\linewidth]{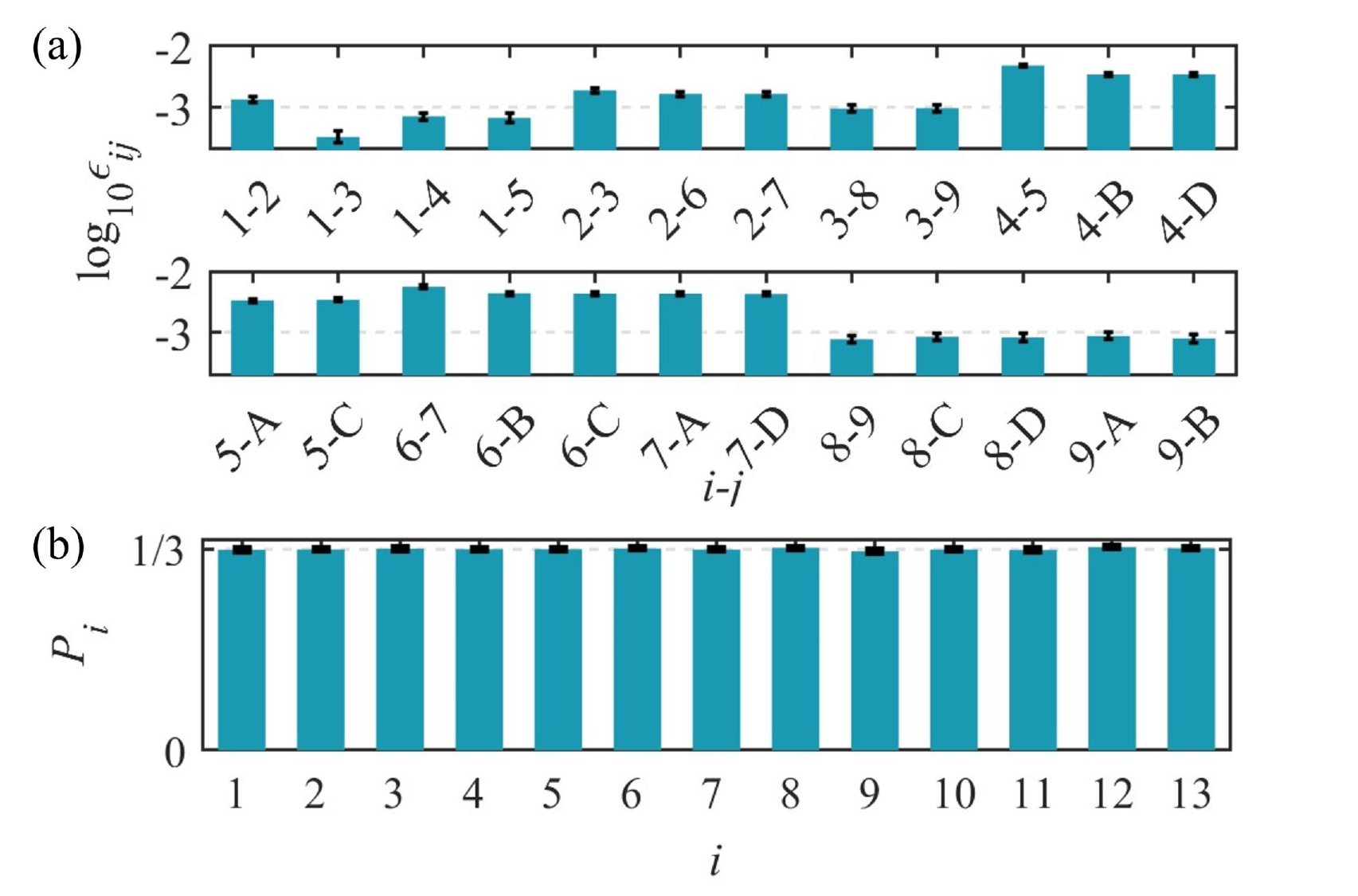}
	\caption{Experiment results of sequentially implementing uncharacterized projectors according to the orthogonality graph of YO-13. $\epsilon_{ij}$ and $P_i$ are shown in (a) and (b), respectively. Experimental error bars are estimated as the standard deviation based on the assumption of photons' Poisson statistics.}
	\label{fig:yo13exp}
\end{figure}


The experiment results of $\epsilon_{ij}$ and $P_i$ are shown in Fig.~\ref{fig:yo13exp}. The values of $\epsilon_{ij}$ are 0 ideally, whose mean value obtained from the experiment equals $0.00232\pm0.00018$. $P_i$ obtained from projection measurement on the maximally mixed state in $d=3$ are $1/3$ ideally, and the mean value of $P_i$ equals $0.3334\pm0.0025$. 

According to the $\epsilon_{ij}'$, $\mathcal W_{\rm worst}$ and $\mathcal W_{\rm SDP}$ are evaluated as $11.661\pm0.028$ and $11.6404\pm0.0043$, which satisfies $\mathcal W_{\rm worst}>\mathcal W_{\rm SDP}$, indicating a successful certification.


\section{Details of evaluating $\delta\theta$}
\label{sec:VII}


Here, we attribute all experimental imperfection of noiseless $\epsilon'_{ij}$ to the calibration angle error $\delta\theta$ of the final wave plate as the representation of imperfections.
According to the Jones matrix of \eqref{eq:jones} and the denotations of HWP angles, we can calculate the ideal angles of wave plates to be set in the experiment. For example, in the M2 block, when projecting the state prepared in P2 onto the state $|v_{24}\rangle=(0,1,-1,0)^T/\sqrt{2}$, the angles of HWP are ideally $\theta^{24}_4=-22.5^\circ$,$\theta^{24}_5=45^\circ$,$\theta^{24}_6=0^\circ$. According to the assumption mentioned above, all angular errors of these three wave plates are attributed to the angle of the last HWP $\theta^j_4+\delta\theta$. Therefore, the state to project $|V_j\rangle$ is not the ideal $|v_j\rangle$ and the projection probability is $|\langle V_j|v_i\rangle|^2$. We perform the optimization program as follows:
\begin{equation}
	\begin{aligned}
		\min_{\delta\theta} \ &\sum_{i,j\in E}(|\langle V_j|v_i\rangle|^2-\epsilon'_{ij})^2,\\
	\end{aligned}
\end{equation}
where $\epsilon'_{ij}$ are the noiseless projection probabilities obtained in the previous section. When the optimum of the optimization program is achieved, we obtain the evaluated $\delta\theta$ from this program to represent the level of imperfection. The results of $\delta\theta$ in each group are shown in Fig.~4 (b).


\bibliography{supmatref}